\documentclass[aps,showpacs,nofootinbib,superscriptaddress,twocolumn,floatfix]{revtex4-1}

\usepackage[utf8]{inputenc}
\usepackage{graphicx}
\usepackage{amssymb}
\usepackage{amsmath}
\usepackage{bm}
\usepackage{slashed}
\usepackage{datetime}
\usepackage{mciteplus}

\begin{document}
\allowdisplaybreaks[2]

\title{Exploring universality of transversity in proton-proton collisions}

\author{Marco Radici}
\email{marco.radici@pv.infn.it}
\affiliation{INFN Sezione di Pavia, via Bassi 6, I-27100 Pavia, Italy}

\author{Alessandro M. Ricci}
\email{alessandromari.ricci01@universitadipavia.it}
\affiliation{Dipartimento di Fisica, Universit\`a di Pavia, via Bassi 6, I-27100 Pavia, Italy}

\author{Alessandro Bacchetta}
\email{alessandro.bacchetta@unipv.it}
\affiliation{Dipartimento di Fisica, Universit\`a di Pavia, via Bassi 6, I-27100 Pavia, Italy}

\author{Asmita Mukherjee}
\email{asmita@phy.iitb.ac.in}
\affiliation{Physics Department, Indian Institute of Technology Bombay, Powai, Mumbai 400076, India}

\begin{abstract}
We consider the azimuthal correlations of charged hadron pairs with large total transverse momentum and small relative momentum, produced in proton-proton collisions with one transversely polarized proton. One of these correlations directly probes the chiral-odd transversity parton distribution in connection with a chiral-odd interference fragmentation function. We present predictions for this observable based on previous extractions of transversity (from charged pion pair production in semi-inclusive deep-inelastic scattering) and of the interference fragmentation function (from the production of back-to-back charged pion pairs in electron-positron annihilations). All analyses are performed in the framework of collinear factorization. We compare our predictions to the recent data on proton-proton collisions released by the {\tt STAR} collaboration at {\tt RHIC}, and we find them reasonably compatible. This comparison  confirms for the first time the predicted role of transversity in proton-proton collisions and it allows to test its universality. 
\end{abstract}

\date{\today, \currenttime}


\maketitle

Parton distribution functions (PDFs) describe combinations of number densities of quarks and gluons in a fast-moving hadron. If the parton transverse momentum is integrated over (collinear framework), the parton structure of spin-$\textstyle{\frac{1}{2}}$ hadrons (like the nucleon) is described at first order in terms of only three PDFs: the unpolarized distribution $f_1$, the longitudinally polarized (helicity) distribution $g_1$, and the transversely polarized (transversity) distribution $h_1$. The $h_1$ is the least known PDF because it is characterized by a parton chirality flip; {\it i.e.}, it is a chiral-odd function. Transversity occurs only in observables where it is paired to a chiral-odd partner. Hence, it can be measured only in processes with two hadrons in the initial state ({\it e.g.}, proton-proton collisions) or one hadron in the initial state and at least one hadron in the final state ({\it e.g.}, semi-inclusive deep-inelastic scattering -- SIDIS). Transversity vanishes for gluons inside the nucleon. Contrary to helicity, under evolution it scales like a pure non-singlet function. Its first Mellin moment, the nucleon tensor charge, belongs to the group of nucleon charges (mostly known only on lattice) that could put constraints on the search of new physics mechanisms beyond the Standard Model~\cite{Cirigliano:2013xha,Bhattacharya:2015esa,Courtoy:2015haa}.   

In this paper, we expose the direct effect of transversity on the distribution of final hadron pairs produced in proton-proton collisions with one transversely polarized proton. We calculate for the first time the corresponding transverse spin asymmetry and we compare our predictions with recent {\tt STAR} data for the case of detected final $(\pi^+ \pi^-)$ pairs at center-of-mass (c.m.) energy $\sqrt{s} = 200$ GeV~\cite{Adamczyk:2015hri}. When the two hadrons are produced with low relative momentum, the asymmetry is proportional to a convolution containing the transversity distribution $h_1$ and its chiral-odd partner denoted $H_1^{\sphericalangle}$~\cite{Bacchetta:2004it}. The $H_1^{\sphericalangle}$ is a specific polarized di-hadron fragmentation function (DiFF) that describes the distortion of the azimuthal distribution of the final hadron pair when the fragmenting parton is transversely polarized~\cite{Bianconi:1999cd}. 

The transversity $h_1$ and the DiFF $H_1^{\sphericalangle}$ were extracted by fitting data for the semi-inclusive production of $(\pi^+ \pi^-)$ pairs in SIDIS and in $e^+ e^-$ annihilations. The cross section for SIDIS at leading twist contains a term which is proportional to the product of $h_1$ and $H_1^{\sphericalangle}$~\cite{Jaffe:1998pv,Radici:2001na,Bacchetta:2002ux}. The same $H_1^{\sphericalangle}$ appears also in the leading-twist cross section for the semi-inclusive back-to-back emission of two hadron pairs from $e^+ e^-$ annihilations~\cite{Artru:1995zu,Boer:2003ya}. 
The $H_1^{\sphericalangle}$ was parametrized~\cite{Courtoy:2012ry} using the $e^+ e^-$ data from the {\tt BELLE} collaboration~\cite{Vossen:2011fk}. The transversity valence components $h_1^{u_v}$ and $h_1^{d_v} $ were extracted~\cite{Bacchetta:2011ip,Bacchetta:2012ty} from the SIDIS data of the {\tt HERMES}~\cite{Airapetian:2008sk} and {\tt COMPASS}~\cite{Adolph:2012nw} collaborations. The analysis has been recently updated~\cite{Radici:2015mwa} by enclosing the latest and more precise {\tt COMPASS} data for a transversely polarized proton target~\cite{Adolph:2014fjw}. 

There is a general consistency (at least, in the kinematical range where there are SIDIS data) between the transversity extracted with the above strategy and the one extracted via the Collins effect in single-hadron SIDIS~\cite{Anselmino:2015sxa,Kang:2015msa} (see also Ref.~\cite{Martin:2014wua}). However, the analysis of the latter case requires a formalism with an explicit dependence on the parton intrinsic momentum, the so-called TMD factorization framework. The TMD framework cannot be applied to single hadron production in hadronic collisions since there are explicit counter-examples showing that TMD factorization is broken in this case~\cite{Rogers:2010dm}. On the contrary, the case of di-hadron production can be analyzed using a collinear factorization framework, which can be applied also in hadronic collisions. This allows us to study transversity in a process different from SIDIS and explore its universality. 

In the following, we take our parametrizations of $h_1$ and $H_1^{\sphericalangle}$, obtained by fitting SIDIS and $e^+ e^-$ data, and we compute the relevant transverse spin azimuthal asymmetry for semi-inclusive $(\pi^+ \pi^-)$ production in proton-proton collisions. Then, we compare our predictions with the {\tt STAR} data. In spite of some limitations in our estimate (in particular, we include only valence quarks because these are the only components of transversity that can be extracted from the current fixed-target SIDIS data), we obtain a reasonable agreement with experimental measurements. This is an important achievement because it gives us confidence in the reliability of the framework and opens a unique opportunity to use hadronic collisions data for the extraction of transversity based on the di-hadron collinear mechanism.

\begin{figure}
\begin{center}
\includegraphics[width=0.5\textwidth]{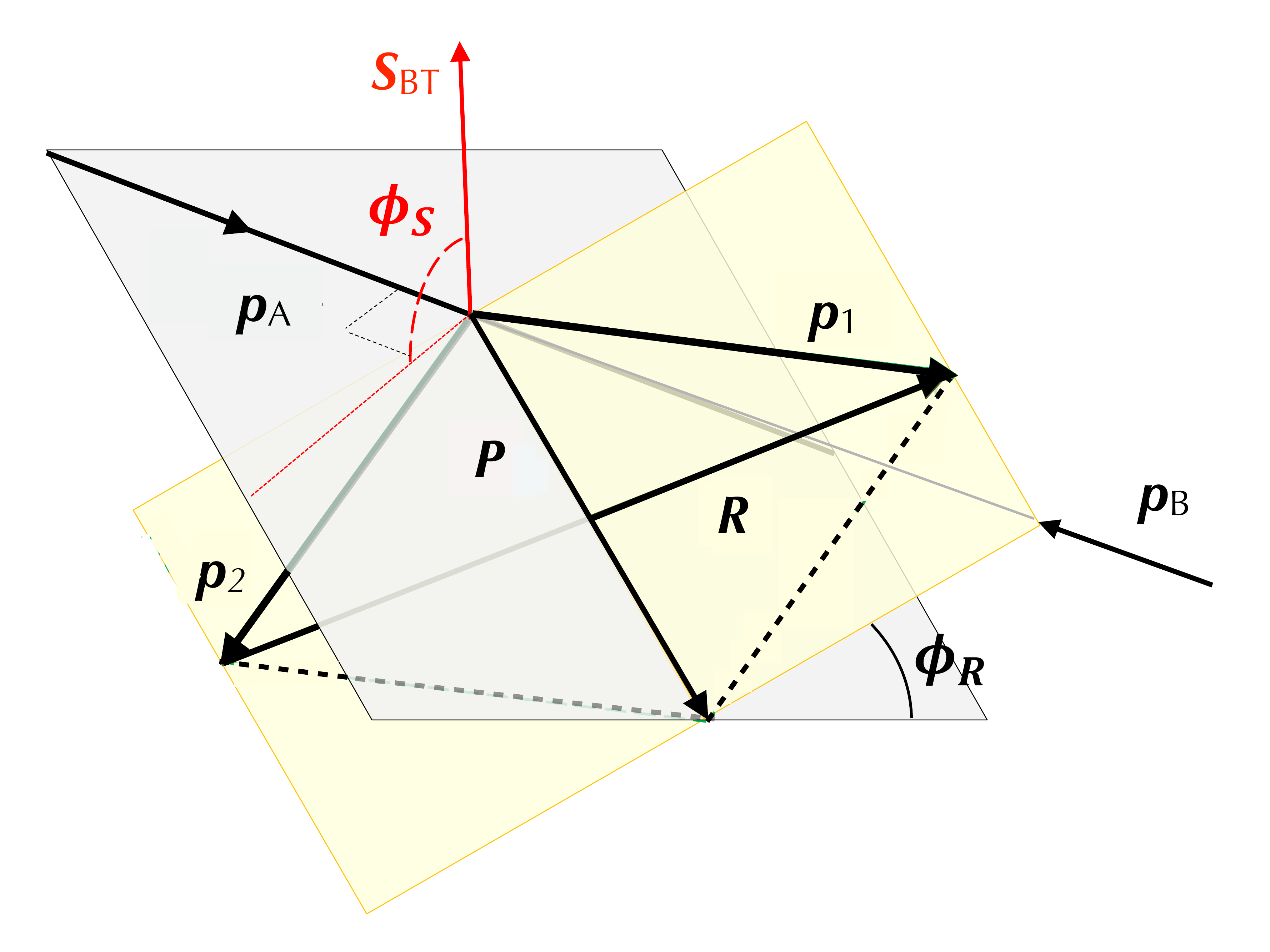}
\end{center}
\caption{\label{fig:kin} Kinematics for the collision of a proton with momentum $\bm{p}_A$ and a tranversely polarized proton with momentum $\bm{p}_B$ and spin vector $\bm{S}_{BT}$. The final state is represented by the inclusive production of two hadrons with total and relative momenta $\bm{P} = \bm{p}_1 + \bm{p}_2$ and $\bm{R} = \bm{p}_1 - \bm{p}_2$, forming a plane oriented with the azimuthal angle $\phi_R^{}$ with respect to the reaction plane formed by $\bm{p}_A$ and $\bm{P}$.}
\end{figure}

We consider the process $p_A + p_B^\uparrow \to (h_1\  h_2) + X$ where a proton with momentum $p_A$ collides with a transversely polarized proton $p_B$ with spin vector $S_B$, and two unpolarized hadrons $h_1$ and $h_2$ (with momenta $p_1$ and $p_2$ and masses $M_1$ and $M_2$, respectively) are inclusively detected inside the same jet. We define the pair total momentum $P = p_1 + p_2$ and relative momentum $R = (p_1 - p_2)/2$; the pair invariant mass is $M_h^2 = P^2$. Since we work in a collinear framework, we integrate over the intrinsic transverse components of $\bm{P}$ with respect to the jet axis. The components of $\bm{P}$ perpendicular to the beam (defined by $\bm{p}_A$) are indicated by $\bm{P}_T$. We identify the reaction plane as the plane formed by $\bm{p}_A$ and $\bm{P}$. All azimuthal angles are measured with respect to this plane (see Fig.~\ref{fig:kin} and Ref.~\cite{Bacchetta:2004it} for a formal definition). The most relevant angles are $\phi_{S}$, the azimuthal angle of the polarization vector $\bm{S}_B$, and $\phi_R$, which describes the azimuthal orientation around $\bm{P}$ of the plane containing the hadron pair momenta $\bm{p}_1$ and $\bm{p}_2$ (see Fig.~\ref{fig:kin}). The modulus $|\bm{P}_T|$ is the hard scale of the process. Hence, we assume that $|\bm{P}_T| \gg M_h, M_1, M_2,$ and we perform our analysis at leading order in $1/|\bm{P}_T|$. The differential cross section reads~\cite{Bacchetta:2004it}
\begin{equation}
\frac{d\sigma}{d\eta \, d|\bm{P}_T| dM_h d\phi^{}_R d\phi_{S}} = d\sigma^0  \left( 1 + \sin (\phi^{}_{S} - \phi^{}_R) A^{}_{UT} \right) , 
\label{e:ppcross}
\end{equation}
where $d\sigma^0$ is the unpolarized cross section 
\begin{align}
\frac{d\sigma^0}{d\eta\, d|\bm{P}_T|\, dM_h} &=  2 \, |\bm{P}_T| \, \sum_{a,b,c,d} \int \frac{d x_a\, dx_b }{4 \pi^2 \bar{z}} \nonumber \\
&\times f_1^a (x_a) \, f_1^b(x_b) \, \frac{d\hat{\sigma}_{ab \to cd}}{d\hat{t}} \, D_1^c (\bar{z}, M_h) \; , 
\label{e:ppcross0}
\end{align}
and the transverse spin asymmetry $A^{}_{UT}$ is given by 
\begin{align}
A^{}_{UT} (\eta, \, &|\bm{P}_T|, \, M_h) = \frac{|\bm{S}_{BT}|\, 2 \, |\bm{P}_T|}{d\sigma^0}\, \frac{|\bm{R}_T|}{M_h}\, \sum_{a,b,c,d}\, \int \frac{dx_a \, dx_b}{16 \pi \bar{z}} \nonumber \\
&\times  f_1^a(x_a) \, h_1^b(x_b) \, \frac{d\Delta \hat{\sigma}_{ab^\uparrow \to c^\uparrow d}}{d\hat{t}} \, H_1^{\sphericalangle c}(\bar{z}, M_h) \, .
\label{e:AUT}
\end{align}

The pseudorapidity $\eta$ of the hadron pair is defined with respect to the beam momentum $\bm{p}_A$. Hence, forward transversely polarized particles are associated to negative pseudorapidities. Experimental data have been presented with the opposite choice~\cite{Adamczyk:2015hri}. In Eq.~\eqref{e:ppcross0}, the elementary cross section $d\hat{\sigma}$ describes the annihilation of partons $a$ and $b$ (carrying fractional momenta $x_a$ and $x_b$, respectively) into the partons $c$ and $d$. The inclusive decay of parton $c$ into the detected hadron pair is described by the unpolarized DiFF $D_1^c$, that depends on the parton fractional energy $z$ carried by the hadron pair and on the invariant mass $M_h$ of the pair itself. Similarly, in Eq.~\eqref{e:AUT} the cross section $d\Delta \hat{\sigma}$ describes the transfer of polarization in the elementary annihilation when parton $b^\uparrow$ is transversely polarized. As previously mentioned, the inclusive fragmentation of the transversely polarized parton $c^\uparrow$ is described by $H_1^{\sphericalangle c}$. From both SIDIS and $e^+ e^-$ data a specific component of $H_1^{\sphericalangle}$ is extracted that corresponds to the $(\pi^+ \pi^-)$ pair being produced in a state with mismatch in relative orbital angular momentum $|\Delta L| = 1$, {\it i.e.} it corresponds to the interference between the amplitudes for the decay into a pair with relative $s$ wave or $p$ wave~\cite{Bacchetta:2002ux}. Accordingly, this component is usually named interference fragmentation function (IFF)~\cite{Jaffe:1998pv}. Since in this context there is no ambiguity, in the following we will keep denoting it as $H_1^{\sphericalangle}$. 

In the above equations, the sum runs upon all possible combinations of parton flavors; the corresponding (polarized) elementary cross sections are listed in the Appendix of Ref.~\cite{Bacchetta:2004it}. Both $d\hat{\sigma}$ and $d\Delta \hat{\sigma}$ are differential in $\hat{t} = t x_a / \bar{z}$, where $t = (p_A - p_B)^2$ and $\bar{z}$ is the value taken by $z$ because of momentum conservation at the partonic level~\cite{Bacchetta:2004it}: 
\begin{equation}
\bar{z} = \frac{|\bm{P}_T|}{\sqrt{s}}\, \frac{x_a e^{-\eta} + x_b e^{\eta}}{x_a x_b} \, , 
\label{e:stu}
\end{equation}
with $s = (p_A + p_B)^2$. Finally, in Eq.~\eqref{e:AUT} $|\bm{R}_T| = |\bm{R}| \sin\theta$, where $\theta$ is the angle between the back-to-back emission direction in the c.m. of the hadron pair and $\bm{P}$ in the reaction plane~\cite{Bacchetta:2002ux}. In the following, $\sin\theta$ will be replaced by the average experimental value in each corresponding bin. The modulus $|\bm{R}|$ is related to the pair invariant mass through~\cite{Bacchetta:2002ux,Bacchetta:2004it}
\begin{equation}
\frac{|\bm{R}|}{M_h} = \frac{1}{2} \, \sqrt{1 - 2 \frac{M_1^2+M_2^2}{M_h^2} + \frac{(M_1^2-M_2^2)^2}{M_h^4}} \; .
\label{e:R}
\end{equation}

Evidence for the transverse spin asymmetry $A^{}_{UT}$ of Eq.~\eqref{e:AUT} has been recently reported by the {\tt STAR} collaboration for the process $p + p^\uparrow \to (\pi^+ \pi^-) + X$ at the c.m. energy $\sqrt{s} = 200$ 
GeV~\cite{Adamczyk:2015hri}. If the final hadron pair is represented by charged pions $(\pi^+ \pi^-)$, then the above formulas simplify because isospin symmetry and charge conjugation induce specific symmetry relations among the various flavor components of 
DiFFs~\cite{Bacchetta:2006un,Bacchetta:2011ip,Bacchetta:2012ty,Radici:2015mwa}. In the following, we compare the experimental results with our predictions for $A^{}_{UT}$. We compute them in two steps. First, we replace in Eq.~\eqref{e:AUT} the DiFFs with those ones used to fit the {\tt BELLE} data on the azimuthal asymmetry of the pion pair distribution in the process $e^+ + e^- \to (\pi^+ \pi^-)_{\mathrm{jet1}} + (\pi^+ \pi^-)_{\mathrm{jet2}} + X$~\cite{Courtoy:2012ry}. Note that the {\tt BELLE} data do not allow to extract parametrizations of $D_1^g$: the gluon contribution is generated only by the effect of QCD evolution. The second step consists in replacing the PDFs with those ones used to fit the {\tt HERMES} and {\tt COMPASS} data on the SIDIS transverse spin asymmetry in $e + p^\uparrow \to e' + (\pi^+ \pi^-) + X$~\cite{Radici:2015mwa}. Specifically, we consider the parametric expressions at $Q_0^2 = 1$ GeV$^2$ of all replicas of DiFFs and transversity that fit the $e^+ e^-$ and SIDIS data, and we evolve each replica to the {\tt STAR} $|\bm{P}_T|$ scales by computing its DGLAP evolution equations at leading order~\cite{Ceccopieri:2007ip}, using the {\tt HOPPET} code~\cite{Salam:2008qg} suitably extended to include chiral-odd splitting functions. The replica method  was systematically applied to the statistical error analysis in the extraction of both DiFFs and transversity. Moreover, we tested different starting expressions for $h_1^{q_v} (x, Q_0^2)$ and two different values of $\alpha_s (M_Z^2)$ in the evolution code, in order to account for the uncertainties in the determination of $\Lambda_{\text{QCD}}$ and, more generally, to include a theoretical systematic error~\cite{Bacchetta:2012ty,Radici:2015mwa}. Thus, we believe that the predictions we are presenting here are based on the current most realistic estimate of the uncertainties on transversity, particularly for kinematical regions outside the range covered by SIDIS experimental data. In the following, we will present results for the choice $\alpha_s (M_Z^2) = 0.139$~\cite{Martin:2009iq} and for the socalled {\it flexible} form of $h_1^{q_v} (x, Q_0^2)$~\cite{Radici:2015mwa}. The other choices behave in a similar way. Moreover, we show our results as an uncertainty band corresponding to the central 68\% of all replicas. 

\begin{figure}[h]
\begin{center}
\includegraphics[width=0.5\textwidth]{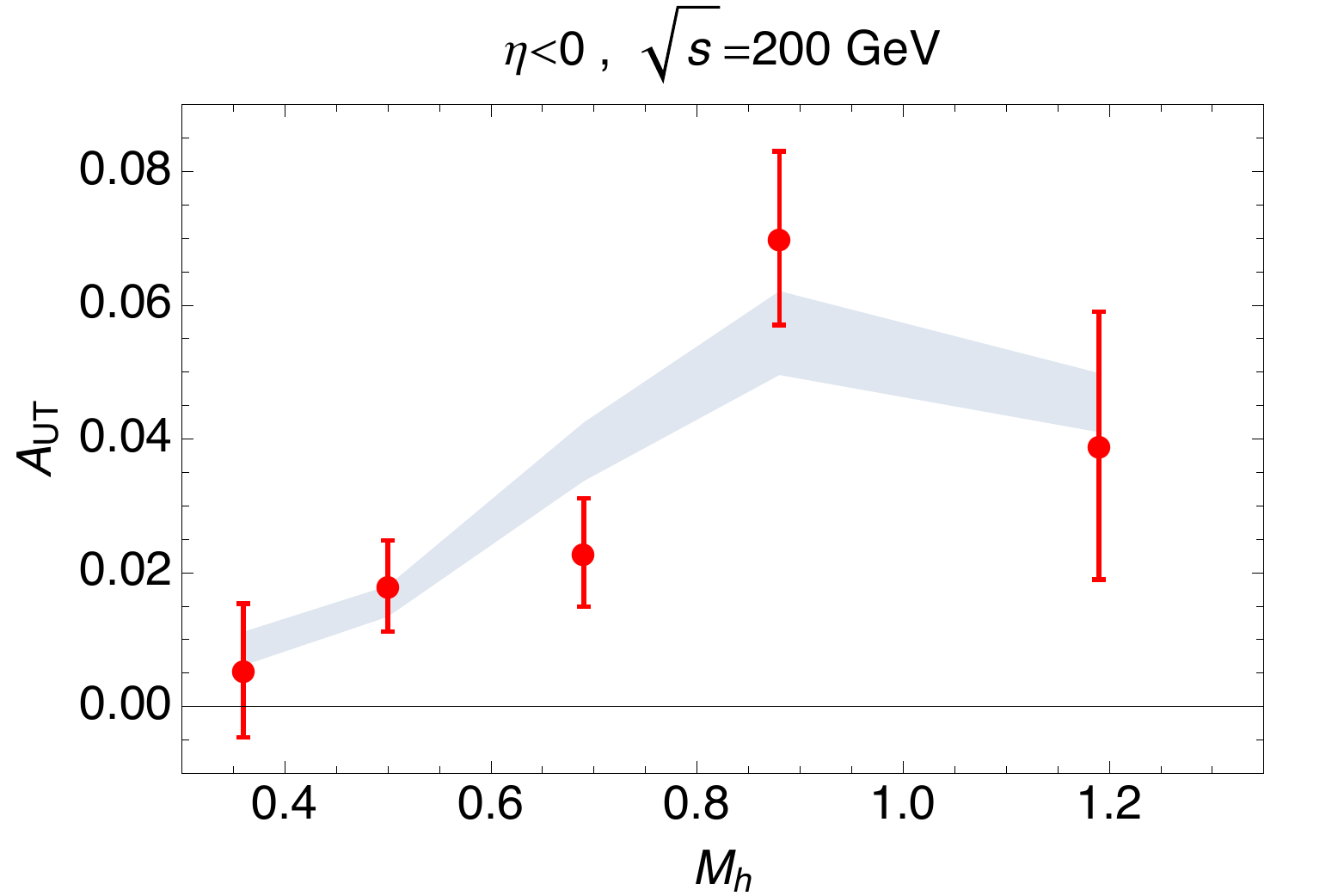}
\end{center}
\caption{\label{fig:AMetaminus} The asymmetry $A^{}_{UT}$ of Eq.~\eqref{e:AUT}  as a function of $M_h$ with $|\bm{P}_T|$ and negative (forward) $\eta$ integrated in all experimental bins. The uncertainty band corresponds to the 68\% of all replicas deduced by fitting the $(\pi^+ \pi^-)$ SIDIS and $e^+ e^-$ data~\cite{Radici:2015mwa} (see text). Data are taken from Ref.~\cite{Adamczyk:2015hri} and adjusted to the conventions of this paper.}
\end{figure}

In Fig.~\ref{fig:AMetaminus}, the transverse spin asymmetry $A^{}_{UT}$ of Eq.~\eqref{e:AUT} is plotted as function of the invariant mass $M_h$, integrating on $|\bm{P}_T|$ and on forward negative $\eta$ in all experimental bins, namely for $3 \leq |\bm{P}_T| \leq 13$ GeV$/c$ and $-1.4 \leq \eta \leq 0$. For each experimental $M_h$ bin, the theoretical result for $A^{}_{UT}$ is deduced by computing the integral on $M_h$ over the width of the bin. The experimental data are the result of the 2006 run performed by the {\tt STAR} collaboration~\cite{Adamczyk:2015hri}. For $\eta < 0$, we have forward-propagating transversely polarized particles: the asymmetry is more sensitive to the contribution at large $x_b$ of valence quarks to transversity and it turns out to be sizeable, as expected. It displays the typical shape of the $M_h$ dependence of DiFFs, namely a bump around the mass of the $\rho$ resonance. The overall agreement between theoretical predictions and data is remarkable. It is confirmed also when looking at $A^{}_{UT}$ as function of $M_h$ at $\eta > 0$. It suggests that the transversity $h_1$ and IFF $H_1^{\sphericalangle}$ could indeed be considered in this first instance as universal functions occurring in the cross sections for the various hard processes leading to the inclusive production of charged pion pairs. 

\begin{figure}
\begin{center}
\includegraphics[width=0.5\textwidth]{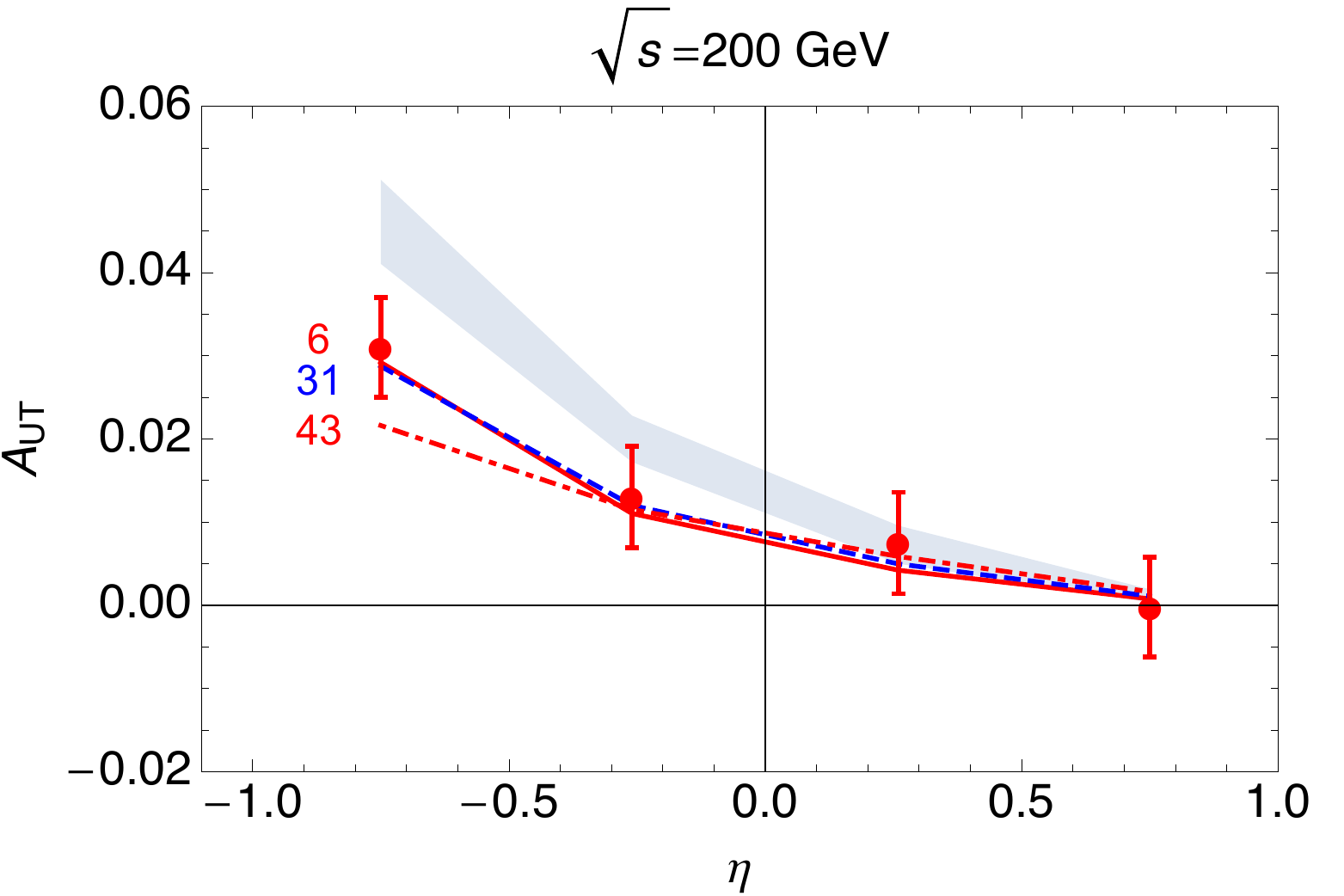}
\end{center}
\caption{\label{fig:Aeta} The asymmetry $A^{}_{UT}$  as a function of pseudorapidity $\eta$, integrated over $M_h$ and $|\bm{P}_T|$. Forward kinematics corresponds to negative $\eta$. Solid (red) line for replica ``6" of $A^{}_{UT}$, dashed  (blue) line for replica ``31", dot-dashed (red) line for replica ``43". Further notation and conventions as in the previous figure.}
\end{figure}

In Fig.~\ref{fig:Aeta}, the asymmetry is shown as function of $\eta$ when integrated on $3 \leq |\bm{P}_T| \leq 13$ GeV$/c$ and $0.3 \leq M_h \leq 1.2$ GeV.\footnote{The indicated $M_h$ range does not overlap with the {\tt STAR} experimental bins at the largest $M_h$~\cite{Adamczyk:2015hri} because the assumptions behind the parametrization of the $M_h$ dependence of DiFFs are valid only up to $M_h \approx 1.2$ GeV~\cite{Bacchetta:2012ty}.} Similarly to the previous figure, for each experimental $\eta$ point the theoretical result is integrated in the corresponding bin. Positive pseudorapidities correspond to backward-propagating transversely polarized particles: the asymmetry is dominated by the contribution of transversely polarized partons with low $x_b$, the transversity is less important, and the resulting asymmetry is largely suppressed. The agreement with data is very good even though the theoretical band is very narrow. This feature is determined by the assumptions adopted in the analysis of $(\pi^+ \pi^-)$ SIDIS data: the low-$x$ behaviour of transversity cannot be fixed yet by the current fixed-target data, and it is imposed by hand to grant that the resulting tensor charge is finite~\cite{Bacchetta:2012ty,Radici:2015mwa}. At $\eta < 0$, the situation is less satisfactory. As explained above, the asymmetry here is larger because it is dominated by the valence components of transversity. Nevertheless, the 68\% band of computed replicas starts to deviate from the experimental points. 

However, we observe that some of the replicas lying outside the 68\% band are close to the data points in this kinematical region. In Fig.~\ref{fig:Aeta}, the solid (red) line refers to the result of replica ``6" for $A^{}_{UT}$, the dashed  (blue) line to replica ``31", and the dot-dashed (red) line to replica ``43". All three results very closely reproduce the experimental points. This is remarkable, if we consider that the curves are predictions. The replicas ``6", ``31", and ``43", do not belong to the 68\% band of replicas that fit the SIDIS data either. But the corresponding $\chi^2$ per degree of freedom in that fit are still reasonably low: 2.04, 1.52, and 2.02, respectively~\cite{Bacchetta:2012ty}. Moreover, they share a very peculiar feature, as it will be clear in the following.

\begin{figure}
\begin{center}
\includegraphics[width=0.5\textwidth]{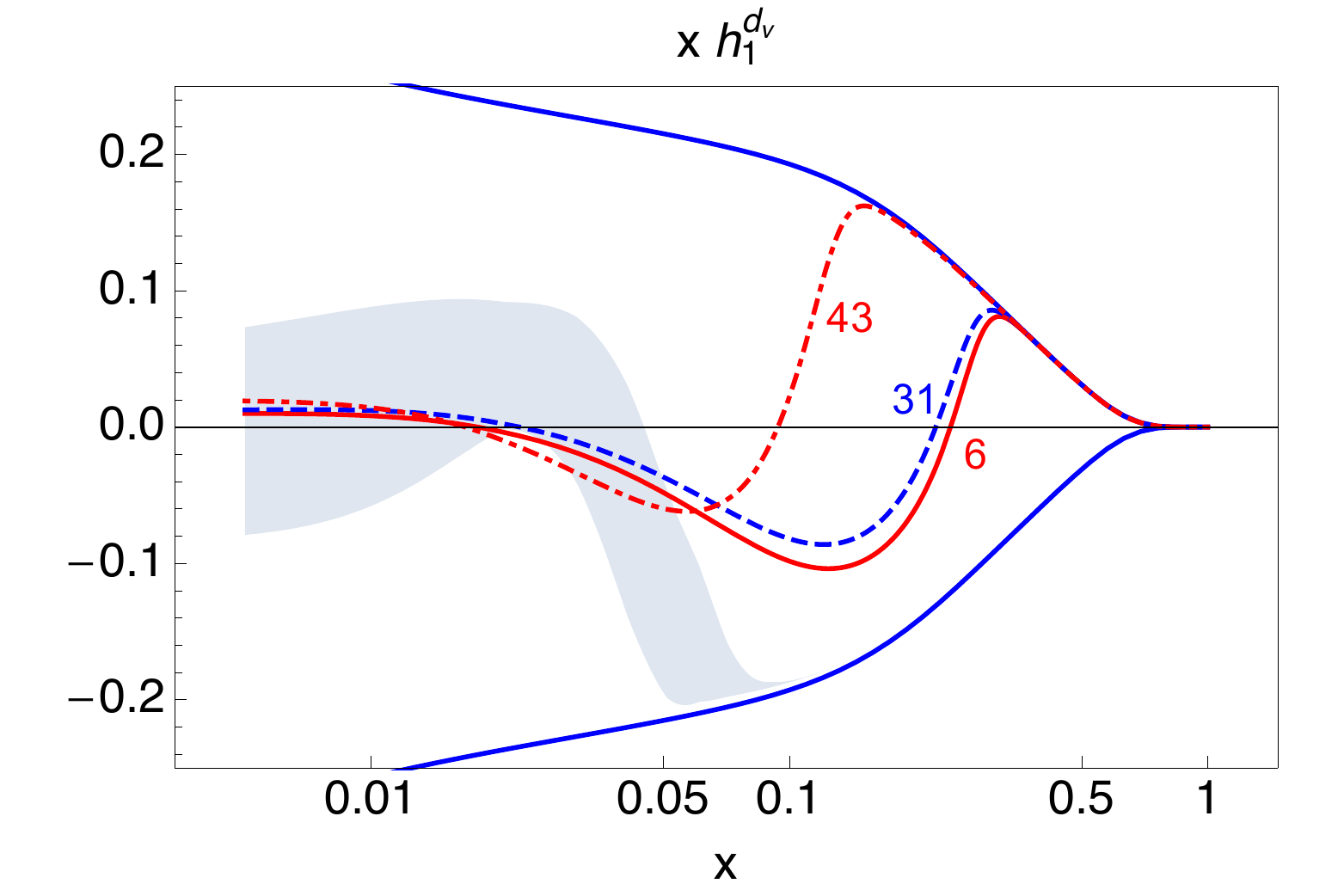}
\end{center}
\caption{\label{fig:xh1dvmark} The valence down transversity $x h_1^{d_v}$ as a function of $x$ at $Q^2 = 2.4$ GeV$^2$. The uncertainty band refers to the 68\% of replicas that fit the $(\pi^+ \pi^-)$ SIDIS data off transversely polarized proton and deuteron targets with the {\it flexible} parametrization and $\alpha_s (M_Z^2) = 0.139$~\cite{Radici:2015mwa}. Dark (blue) solid lines with no label for the upper and lower limits of the Soffer bound. Light (red) solid line for the transversity from replica ``6", dashed (blue) line for replica ``31", dot-dashed (red) line for replica ``43".}
\end{figure}

In Fig.~\ref{fig:xh1dvmark}, we show the uncertainty band for the 68\% of all replicas of the valence down transversity $x h_1^{d_v}$ as a function of $x$ at $Q^2 = 2.4$ GeV$^2$, that fit the SIDIS data for semi-inclusive production of $(\pi^+ \pi^-)$ pairs on transversely polarized proton and deuteron targets. The dark solid lines with no labels represent the upper and lower limits of the Soffer bound. The plot corresponds to the darker band with solid borders in the right panel of Fig.~8 in Ref.~\cite{Radici:2015mwa}. The replicas in the band tend to saturate the lower limit of the Soffer bound because they are driven by the {\tt COMPASS} deuteron data, in particular by the 7th and 8th bins in Ref.~\cite{Radici:2015mwa}. The light (red) solid line with label ``6" reproduces the transversity from the corresponding replica. Similarly, the dashed (blue) line refers to replica ``31", while the dot-dashed (red) line to replica ``43". 

Their trajectories do not follow the trend of the 68\% band at large $x$. Rather, they deviate towards the upper Soffer bound and they saturate it. Hence, at large $x\gtrsim 0.1$ there seems to be a tension between the {\tt COMPASS} deuteron data and the {\tt STAR} data for forward asymmetries. However, this tension concerns at most two bins in $x$ of the SIDIS analysis. It is reasonable to expect that when more data for semi-inclusive $(\pi^+ \pi^-)$ production will become available (particularly from the analysis of runs 2011 and 2012 for proton-proton collisions at c.m. energies of $\sqrt{500}$ and $\sqrt{200}$ GeV, respectively, by the {\tt STAR} collaboration) and a global fit of SIDIS, $e^+ e^-$, and proton-proton collisions will become feasible, the above two bins in $x$ from the SIDIS deuteron data will become statistically less relevant and the above tension will most likely disappear. 

In summary, for the first time we have exposed the predicted connection between the experimental evidence of transverse-spin azimuthal asymmetries in the distribution of $(\pi^+ \pi^-)$ pairs produced from proton-proton collisions, and the transversity parton distribution that was extracted from a previous analysis of pion pair production in semi-inclusive deep-inelastic scattering. By combining transversity with its chiral-odd partner, the interference fragmentation function independently extracted from production of back-to-back pion pairs in electron-positron annihilations, we have made predictions for the distribution of $(\pi^+ \pi^-)$ pairs produced in proton-proton collisions with one transversely polarized proton. These calculations can be meaningfully performed in the collinear framework, where the cross sections for the above hard processes can be expressed in a factorized form. By comparing our predictions to the recently released data by the {\tt STAR} collaboration at {\tt RHIC}, we deduce that there are clear and encouraging hints that the involved parton densities are universal, although we encountered some discrepancies in the case of forward kinematics. Further insight into these discrepancies could be gained by collecting more experimental data. For example, it would be important to improve our knowledge of the gluon contribution to the $(\pi^+ \pi^-)$ fragmentation, which is not constrained by the $e^+ e^-$ data. To this aim, measurements of the unpolarized cross section for the $p + p \to (\pi^+ \pi^-) + X$ process would be very useful. The published set of data points for the transverse-spin asymmetry is small and with relatively limited statistics. In the near future, more data are expected from the {\tt STAR} collaboration that will also make it feasible to extract transversity from a global fit of data from all hard processes for the semi-inclusive production of pion pairs.


We thank the {\tt STAR} collaboration and, in particular, Anselm Vossen for many fruitful 
discussions. This research is partially supported by the European Research Council (ERC) under the European 
Union's Horizon 2020 research and innovation programme (grant agreement No. 647981, 3DSPIN). 


\bibliographystyle{apsrevM}
\bibliography{mybiblio}

\ifx\mcitethebibliography\mciteundefinedmacro
\PackageError{apsrevM.bst}{mciteplus.sty has not been loaded}
{This bibstyle requires the use of the mciteplus package.}\fi
\begin{mcitethebibliography}{27}
\expandafter\ifx\csname natexlab\endcsname\relax\def\natexlab#1{#1}\fi
\expandafter\ifx\csname bibnamefont\endcsname\relax
  \def\bibnamefont#1{#1}\fi
\expandafter\ifx\csname bibfnamefont\endcsname\relax
  \def\bibfnamefont#1{#1}\fi
\expandafter\ifx\csname citenamefont\endcsname\relax
  \def\citenamefont#1{#1}\fi
\expandafter\ifx\csname url\endcsname\relax
  \def\url#1{\texttt{#1}}\fi
\expandafter\ifx\csname urlprefix\endcsname\relax\def\urlprefix{URL }\fi
\providecommand{\bibinfo}[2]{#2}
\providecommand{\eprint}[2][]{\url{#2}}

\bibitem[{\citenamefont{Cirigliano et~al.}(2013)\citenamefont{Cirigliano,
  Gardner, and Holstein}}]{Cirigliano:2013xha}
\bibinfo{author}{\bibfnamefont{V.}~\bibnamefont{Cirigliano}},
  \bibinfo{author}{\bibfnamefont{S.}~\bibnamefont{Gardner}}, \bibnamefont{and}
  \bibinfo{author}{\bibfnamefont{B.}~\bibnamefont{Holstein}},
  \bibinfo{journal}{Prog.Part.Nucl.Phys.} \textbf{\bibinfo{volume}{71}},
  \bibinfo{pages}{93} (\bibinfo{year}{2013}), \eprint{1303.6953}\relax
\mciteBstWouldAddEndPuncttrue
\mciteSetBstMidEndSepPunct{\mcitedefaultmidpunct}
{\mcitedefaultendpunct}{\mcitedefaultseppunct}\relax
\EndOfBibitem
\bibitem[{\citenamefont{Bhattacharya et~al.}(2015)\citenamefont{Bhattacharya,
  Cirigliano, Gupta, Lin, and Yoon}}]{Bhattacharya:2015esa}
\bibinfo{author}{\bibfnamefont{T.}~\bibnamefont{Bhattacharya}},
  \bibinfo{author}{\bibfnamefont{V.}~\bibnamefont{Cirigliano}},
  \bibinfo{author}{\bibfnamefont{R.}~\bibnamefont{Gupta}},
  \bibinfo{author}{\bibfnamefont{H.-W.} \bibnamefont{Lin}}, \bibnamefont{and}
  \bibinfo{author}{\bibfnamefont{B.}~\bibnamefont{Yoon}},
  \bibinfo{journal}{Phys. Rev. Lett.} \textbf{\bibinfo{volume}{115}},
  \bibinfo{pages}{212002} (\bibinfo{year}{2015}), \eprint{1506.04196}\relax
\mciteBstWouldAddEndPuncttrue
\mciteSetBstMidEndSepPunct{\mcitedefaultmidpunct}
{\mcitedefaultendpunct}{\mcitedefaultseppunct}\relax
\EndOfBibitem
\bibitem[{\citenamefont{Courtoy et~al.}(2015)\citenamefont{Courtoy, Baessler,
  Gonzalez-Alonso, and Liuti}}]{Courtoy:2015haa}
\bibinfo{author}{\bibfnamefont{A.}~\bibnamefont{Courtoy}},
  \bibinfo{author}{\bibfnamefont{S.}~\bibnamefont{Baessler}},
  \bibinfo{author}{\bibfnamefont{M.}~\bibnamefont{Gonzalez-Alonso}},
  \bibnamefont{and} \bibinfo{author}{\bibfnamefont{S.}~\bibnamefont{Liuti}},
  \bibinfo{journal}{Phys. Rev. Lett.} \textbf{\bibinfo{volume}{115}},
  \bibinfo{pages}{162001} (\bibinfo{year}{2015}), \eprint{1503.06814}\relax
\mciteBstWouldAddEndPuncttrue
\mciteSetBstMidEndSepPunct{\mcitedefaultmidpunct}
{\mcitedefaultendpunct}{\mcitedefaultseppunct}\relax
\EndOfBibitem
\bibitem[{\citenamefont{Adamczyk et~al.}(2015)}]{Adamczyk:2015hri}
\bibinfo{author}{\bibfnamefont{L.}~\bibnamefont{Adamczyk}} \bibnamefont{et~al.}
  (\bibinfo{collaboration}{STAR}), \bibinfo{journal}{Phys. Rev. Lett.}
  \textbf{\bibinfo{volume}{115}}, \bibinfo{pages}{242501}
  (\bibinfo{year}{2015}), \eprint{1504.00415}\relax
\mciteBstWouldAddEndPuncttrue
\mciteSetBstMidEndSepPunct{\mcitedefaultmidpunct}
{\mcitedefaultendpunct}{\mcitedefaultseppunct}\relax
\EndOfBibitem
\bibitem[{\citenamefont{Bacchetta and Radici}(2004)}]{Bacchetta:2004it}
\bibinfo{author}{\bibfnamefont{A.}~\bibnamefont{Bacchetta}} \bibnamefont{and}
  \bibinfo{author}{\bibfnamefont{M.}~\bibnamefont{Radici}},
  \bibinfo{journal}{Phys. Rev.} \textbf{\bibinfo{volume}{D70}},
  \bibinfo{pages}{094032} (\bibinfo{year}{2004}), \eprint{hep-ph/0409174}\relax
\mciteBstWouldAddEndPuncttrue
\mciteSetBstMidEndSepPunct{\mcitedefaultmidpunct}
{\mcitedefaultendpunct}{\mcitedefaultseppunct}\relax
\EndOfBibitem
\bibitem[{\citenamefont{Bianconi et~al.}(2000)\citenamefont{Bianconi, Boffi,
  Jakob, and Radici}}]{Bianconi:1999cd}
\bibinfo{author}{\bibfnamefont{A.}~\bibnamefont{Bianconi}},
  \bibinfo{author}{\bibfnamefont{S.}~\bibnamefont{Boffi}},
  \bibinfo{author}{\bibfnamefont{R.}~\bibnamefont{Jakob}}, \bibnamefont{and}
  \bibinfo{author}{\bibfnamefont{M.}~\bibnamefont{Radici}},
  \bibinfo{journal}{Phys. Rev.} \textbf{\bibinfo{volume}{D62}},
  \bibinfo{pages}{034008} (\bibinfo{year}{2000}),
  \eprint[http://arXiv.org/abs]{hep-ph/9907475}\relax
\mciteBstWouldAddEndPuncttrue
\mciteSetBstMidEndSepPunct{\mcitedefaultmidpunct}
{\mcitedefaultendpunct}{\mcitedefaultseppunct}\relax
\EndOfBibitem
\bibitem[{\citenamefont{Jaffe et~al.}(1998)\citenamefont{Jaffe, Jin, and
  Tang}}]{Jaffe:1998pv}
\bibinfo{author}{\bibfnamefont{R.~L.} \bibnamefont{Jaffe}},
  \bibinfo{author}{\bibfnamefont{X.}~\bibnamefont{Jin}}, \bibnamefont{and}
  \bibinfo{author}{\bibfnamefont{J.}~\bibnamefont{Tang}},
  \bibinfo{journal}{Phys. Rev.} \textbf{\bibinfo{volume}{D57}},
  \bibinfo{pages}{5920} (\bibinfo{year}{1998}),
  \eprint[http://arXiv.org/abs]{hep-ph/9710561}\relax
\mciteBstWouldAddEndPuncttrue
\mciteSetBstMidEndSepPunct{\mcitedefaultmidpunct}
{\mcitedefaultendpunct}{\mcitedefaultseppunct}\relax
\EndOfBibitem
\bibitem[{\citenamefont{Radici et~al.}(2002)\citenamefont{Radici, Jakob, and
  Bianconi}}]{Radici:2001na}
\bibinfo{author}{\bibfnamefont{M.}~\bibnamefont{Radici}},
  \bibinfo{author}{\bibfnamefont{R.}~\bibnamefont{Jakob}}, \bibnamefont{and}
  \bibinfo{author}{\bibfnamefont{A.}~\bibnamefont{Bianconi}},
  \bibinfo{journal}{Phys. Rev.} \textbf{\bibinfo{volume}{D65}},
  \bibinfo{pages}{074031} (\bibinfo{year}{2002}),
  \eprint[http://arXiv.org/abs]{hep-ph/0110252}\relax
\mciteBstWouldAddEndPuncttrue
\mciteSetBstMidEndSepPunct{\mcitedefaultmidpunct}
{\mcitedefaultendpunct}{\mcitedefaultseppunct}\relax
\EndOfBibitem
\bibitem[{\citenamefont{Bacchetta and Radici}(2003)}]{Bacchetta:2002ux}
\bibinfo{author}{\bibfnamefont{A.}~\bibnamefont{Bacchetta}} \bibnamefont{and}
  \bibinfo{author}{\bibfnamefont{M.}~\bibnamefont{Radici}},
  \bibinfo{journal}{Phys. Rev.} \textbf{\bibinfo{volume}{D67}},
  \bibinfo{pages}{094002} (\bibinfo{year}{2003}), \eprint{hep-ph/0212300}\relax
\mciteBstWouldAddEndPuncttrue
\mciteSetBstMidEndSepPunct{\mcitedefaultmidpunct}
{\mcitedefaultendpunct}{\mcitedefaultseppunct}\relax
\EndOfBibitem
\bibitem[{\citenamefont{Artru and Collins}(1996)}]{Artru:1995zu}
\bibinfo{author}{\bibfnamefont{X.}~\bibnamefont{Artru}} \bibnamefont{and}
  \bibinfo{author}{\bibfnamefont{J.~C.} \bibnamefont{Collins}},
  \bibinfo{journal}{Z.Phys.} \textbf{\bibinfo{volume}{C69}},
  \bibinfo{pages}{277} (\bibinfo{year}{1996}), \eprint{hep-ph/9504220}\relax
\mciteBstWouldAddEndPuncttrue
\mciteSetBstMidEndSepPunct{\mcitedefaultmidpunct}
{\mcitedefaultendpunct}{\mcitedefaultseppunct}\relax
\EndOfBibitem
\bibitem[{\citenamefont{Boer et~al.}(2003)\citenamefont{Boer, Jakob, and
  Radici}}]{Boer:2003ya}
\bibinfo{author}{\bibfnamefont{D.}~\bibnamefont{Boer}},
  \bibinfo{author}{\bibfnamefont{R.}~\bibnamefont{Jakob}}, \bibnamefont{and}
  \bibinfo{author}{\bibfnamefont{M.}~\bibnamefont{Radici}},
  \bibinfo{journal}{Phys. Rev.} \textbf{\bibinfo{volume}{D67}},
  \bibinfo{pages}{094003} (\bibinfo{year}{2003}), \eprint{hep-ph/0302232}\relax
\mciteBstWouldAddEndPuncttrue
\mciteSetBstMidEndSepPunct{\mcitedefaultmidpunct}
{\mcitedefaultendpunct}{\mcitedefaultseppunct}\relax
\EndOfBibitem
\bibitem[{\citenamefont{Courtoy et~al.}(2012)\citenamefont{Courtoy, Bacchetta,
  Radici, and Bianconi}}]{Courtoy:2012ry}
\bibinfo{author}{\bibfnamefont{A.}~\bibnamefont{Courtoy}},
  \bibinfo{author}{\bibfnamefont{A.}~\bibnamefont{Bacchetta}},
  \bibinfo{author}{\bibfnamefont{M.}~\bibnamefont{Radici}}, \bibnamefont{and}
  \bibinfo{author}{\bibfnamefont{A.}~\bibnamefont{Bianconi}},
  \bibinfo{journal}{Phys.Rev.} \textbf{\bibinfo{volume}{D85}},
  \bibinfo{pages}{114023} (\bibinfo{year}{2012}), \eprint{1202.0323}\relax
\mciteBstWouldAddEndPuncttrue
\mciteSetBstMidEndSepPunct{\mcitedefaultmidpunct}
{\mcitedefaultendpunct}{\mcitedefaultseppunct}\relax
\EndOfBibitem
\bibitem[{\citenamefont{Vossen et~al.}(2011)}]{Vossen:2011fk}
\bibinfo{author}{\bibfnamefont{A.}~\bibnamefont{Vossen}} \bibnamefont{et~al.}
  (\bibinfo{collaboration}{Belle Collaboration}),
  \bibinfo{journal}{Phys.Rev.Lett.} \textbf{\bibinfo{volume}{107}},
  \bibinfo{pages}{072004} (\bibinfo{year}{2011}), \eprint{1104.2425}\relax
\mciteBstWouldAddEndPuncttrue
\mciteSetBstMidEndSepPunct{\mcitedefaultmidpunct}
{\mcitedefaultendpunct}{\mcitedefaultseppunct}\relax
\EndOfBibitem
\bibitem[{\citenamefont{Bacchetta et~al.}(2011)\citenamefont{Bacchetta,
  Courtoy, and Radici}}]{Bacchetta:2011ip}
\bibinfo{author}{\bibfnamefont{A.}~\bibnamefont{Bacchetta}},
  \bibinfo{author}{\bibfnamefont{A.}~\bibnamefont{Courtoy}}, \bibnamefont{and}
  \bibinfo{author}{\bibfnamefont{M.}~\bibnamefont{Radici}},
  \bibinfo{journal}{Phys.Rev.Lett.} \textbf{\bibinfo{volume}{107}},
  \bibinfo{pages}{012001} (\bibinfo{year}{2011}), \eprint{1104.3855}\relax
\mciteBstWouldAddEndPuncttrue
\mciteSetBstMidEndSepPunct{\mcitedefaultmidpunct}
{\mcitedefaultendpunct}{\mcitedefaultseppunct}\relax
\EndOfBibitem
\bibitem[{\citenamefont{Bacchetta et~al.}(2013)\citenamefont{Bacchetta,
  Courtoy, and Radici}}]{Bacchetta:2012ty}
\bibinfo{author}{\bibfnamefont{A.}~\bibnamefont{Bacchetta}},
  \bibinfo{author}{\bibfnamefont{A.}~\bibnamefont{Courtoy}}, \bibnamefont{and}
  \bibinfo{author}{\bibfnamefont{M.}~\bibnamefont{Radici}},
  \bibinfo{journal}{JHEP} \textbf{\bibinfo{volume}{1303}}, \bibinfo{pages}{119}
  (\bibinfo{year}{2013}), \eprint{1212.3568}\relax
\mciteBstWouldAddEndPuncttrue
\mciteSetBstMidEndSepPunct{\mcitedefaultmidpunct}
{\mcitedefaultendpunct}{\mcitedefaultseppunct}\relax
\EndOfBibitem
\bibitem[{\citenamefont{Airapetian et~al.}(2008)}]{Airapetian:2008sk}
\bibinfo{author}{\bibfnamefont{A.}~\bibnamefont{Airapetian}}
  \bibnamefont{et~al.} (\bibinfo{collaboration}{HERMES}),
  \bibinfo{journal}{JHEP} \textbf{\bibinfo{volume}{06}}, \bibinfo{pages}{017}
  (\bibinfo{year}{2008}), \eprint{0803.2367}\relax
\mciteBstWouldAddEndPuncttrue
\mciteSetBstMidEndSepPunct{\mcitedefaultmidpunct}
{\mcitedefaultendpunct}{\mcitedefaultseppunct}\relax
\EndOfBibitem
\bibitem[{\citenamefont{Adolph et~al.}(2012)}]{Adolph:2012nw}
\bibinfo{author}{\bibfnamefont{C.}~\bibnamefont{Adolph}} \bibnamefont{et~al.}
  (\bibinfo{collaboration}{COMPASS}), \bibinfo{journal}{Phys.Lett.}
  \textbf{\bibinfo{volume}{B713}}, \bibinfo{pages}{10} (\bibinfo{year}{2012}),
  \eprint{1202.6150}\relax
\mciteBstWouldAddEndPuncttrue
\mciteSetBstMidEndSepPunct{\mcitedefaultmidpunct}
{\mcitedefaultendpunct}{\mcitedefaultseppunct}\relax
\EndOfBibitem
\bibitem[{\citenamefont{Radici et~al.}(2015)\citenamefont{Radici, Courtoy,
  Bacchetta, and Guagnelli}}]{Radici:2015mwa}
\bibinfo{author}{\bibfnamefont{M.}~\bibnamefont{Radici}},
  \bibinfo{author}{\bibfnamefont{A.}~\bibnamefont{Courtoy}},
  \bibinfo{author}{\bibfnamefont{A.}~\bibnamefont{Bacchetta}},
  \bibnamefont{and}
  \bibinfo{author}{\bibfnamefont{M.}~\bibnamefont{Guagnelli}},
  \bibinfo{journal}{JHEP} \textbf{\bibinfo{volume}{05}}, \bibinfo{pages}{123}
  (\bibinfo{year}{2015}), \eprint{1503.03495}\relax
\mciteBstWouldAddEndPuncttrue
\mciteSetBstMidEndSepPunct{\mcitedefaultmidpunct}
{\mcitedefaultendpunct}{\mcitedefaultseppunct}\relax
\EndOfBibitem
\bibitem[{\citenamefont{Adolph et~al.}(2014)}]{Adolph:2014fjw}
\bibinfo{author}{\bibfnamefont{C.}~\bibnamefont{Adolph}} \bibnamefont{et~al.}
  (\bibinfo{collaboration}{COMPASS}), \bibinfo{journal}{Phys.Lett.}
  \textbf{\bibinfo{volume}{B736}}, \bibinfo{pages}{124} (\bibinfo{year}{2014}),
  \eprint{1401.7873}\relax
\mciteBstWouldAddEndPuncttrue
\mciteSetBstMidEndSepPunct{\mcitedefaultmidpunct}
{\mcitedefaultendpunct}{\mcitedefaultseppunct}\relax
\EndOfBibitem
\bibitem[{\citenamefont{Anselmino et~al.}(2015)\citenamefont{Anselmino,
  Boglione, D'Alesio, Gonzalez~Hernandez, Melis, Murgia, and
  Prokudin}}]{Anselmino:2015sxa}
\bibinfo{author}{\bibfnamefont{M.}~\bibnamefont{Anselmino}},
  \bibinfo{author}{\bibfnamefont{M.}~\bibnamefont{Boglione}},
  \bibinfo{author}{\bibfnamefont{U.}~\bibnamefont{D'Alesio}},
  \bibinfo{author}{\bibfnamefont{J.~O.} \bibnamefont{Gonzalez~Hernandez}},
  \bibinfo{author}{\bibfnamefont{S.}~\bibnamefont{Melis}},
  \bibinfo{author}{\bibfnamefont{F.}~\bibnamefont{Murgia}}, \bibnamefont{and}
  \bibinfo{author}{\bibfnamefont{A.}~\bibnamefont{Prokudin}},
  \bibinfo{journal}{Phys. Rev.} \textbf{\bibinfo{volume}{D92}},
  \bibinfo{pages}{114023} (\bibinfo{year}{2015}), \eprint{1510.05389}\relax
\mciteBstWouldAddEndPuncttrue
\mciteSetBstMidEndSepPunct{\mcitedefaultmidpunct}
{\mcitedefaultendpunct}{\mcitedefaultseppunct}\relax
\EndOfBibitem
\bibitem[{\citenamefont{Kang et~al.}(2016)\citenamefont{Kang, Prokudin, Sun,
  and Yuan}}]{Kang:2015msa}
\bibinfo{author}{\bibfnamefont{Z.-B.} \bibnamefont{Kang}},
  \bibinfo{author}{\bibfnamefont{A.}~\bibnamefont{Prokudin}},
  \bibinfo{author}{\bibfnamefont{P.}~\bibnamefont{Sun}}, \bibnamefont{and}
  \bibinfo{author}{\bibfnamefont{F.}~\bibnamefont{Yuan}},
  \bibinfo{journal}{Phys. Rev.} \textbf{\bibinfo{volume}{D93}},
  \bibinfo{pages}{014009} (\bibinfo{year}{2016}), \eprint{1505.05589}\relax
\mciteBstWouldAddEndPuncttrue
\mciteSetBstMidEndSepPunct{\mcitedefaultmidpunct}
{\mcitedefaultendpunct}{\mcitedefaultseppunct}\relax
\EndOfBibitem
\bibitem[{\citenamefont{Martin et~al.}(2015)\citenamefont{Martin, Bradamante,
  and Barone}}]{Martin:2014wua}
\bibinfo{author}{\bibfnamefont{A.}~\bibnamefont{Martin}},
  \bibinfo{author}{\bibfnamefont{F.}~\bibnamefont{Bradamante}},
  \bibnamefont{and} \bibinfo{author}{\bibfnamefont{V.}~\bibnamefont{Barone}},
  \bibinfo{journal}{Phys.Rev.} \textbf{\bibinfo{volume}{D91}},
  \bibinfo{pages}{014034} (\bibinfo{year}{2015}), \eprint{1412.5946}\relax
\mciteBstWouldAddEndPuncttrue
\mciteSetBstMidEndSepPunct{\mcitedefaultmidpunct}
{\mcitedefaultendpunct}{\mcitedefaultseppunct}\relax
\EndOfBibitem
\bibitem[{\citenamefont{Rogers and Mulders}(2010)}]{Rogers:2010dm}
\bibinfo{author}{\bibfnamefont{T.~C.} \bibnamefont{Rogers}} \bibnamefont{and}
  \bibinfo{author}{\bibfnamefont{P.~J.} \bibnamefont{Mulders}},
  \bibinfo{journal}{Phys. Rev.} \textbf{\bibinfo{volume}{D81}},
  \bibinfo{pages}{094006} (\bibinfo{year}{2010}), \eprint{1001.2977}\relax
\mciteBstWouldAddEndPuncttrue
\mciteSetBstMidEndSepPunct{\mcitedefaultmidpunct}
{\mcitedefaultendpunct}{\mcitedefaultseppunct}\relax
\EndOfBibitem
\bibitem[{\citenamefont{Bacchetta and Radici}(2006)}]{Bacchetta:2006un}
\bibinfo{author}{\bibfnamefont{A.}~\bibnamefont{Bacchetta}} \bibnamefont{and}
  \bibinfo{author}{\bibfnamefont{M.}~\bibnamefont{Radici}},
  \bibinfo{journal}{Phys. Rev.} \textbf{\bibinfo{volume}{D74}},
  \bibinfo{pages}{114007} (\bibinfo{year}{2006}), \eprint{hep-ph/0608037}\relax
\mciteBstWouldAddEndPuncttrue
\mciteSetBstMidEndSepPunct{\mcitedefaultmidpunct}
{\mcitedefaultendpunct}{\mcitedefaultseppunct}\relax
\EndOfBibitem
\bibitem[{\citenamefont{Ceccopieri et~al.}(2007)\citenamefont{Ceccopieri,
  Radici, and Bacchetta}}]{Ceccopieri:2007ip}
\bibinfo{author}{\bibfnamefont{F.~A.} \bibnamefont{Ceccopieri}},
  \bibinfo{author}{\bibfnamefont{M.}~\bibnamefont{Radici}}, \bibnamefont{and}
  \bibinfo{author}{\bibfnamefont{A.}~\bibnamefont{Bacchetta}},
  \bibinfo{journal}{Phys. Lett.} \textbf{\bibinfo{volume}{B650}},
  \bibinfo{pages}{81} (\bibinfo{year}{2007}), \eprint{hep-ph/0703265}\relax
\mciteBstWouldAddEndPuncttrue
\mciteSetBstMidEndSepPunct{\mcitedefaultmidpunct}
{\mcitedefaultendpunct}{\mcitedefaultseppunct}\relax
\EndOfBibitem
\bibitem[{\citenamefont{Salam and Rojo}(2009)}]{Salam:2008qg}
\bibinfo{author}{\bibfnamefont{G.~P.} \bibnamefont{Salam}} \bibnamefont{and}
  \bibinfo{author}{\bibfnamefont{J.}~\bibnamefont{Rojo}},
  \bibinfo{journal}{Comput.Phys.Commun.} \textbf{\bibinfo{volume}{180}},
  \bibinfo{pages}{120} (\bibinfo{year}{2009}), \eprint{0804.3755}\relax
\mciteBstWouldAddEndPuncttrue
\mciteSetBstMidEndSepPunct{\mcitedefaultmidpunct}
{\mcitedefaultendpunct}{\mcitedefaultseppunct}\relax
\EndOfBibitem
\bibitem[{\citenamefont{Martin et~al.}(2009)\citenamefont{Martin, Stirling,
  Thorne, and Watt}}]{Martin:2009iq}
\bibinfo{author}{\bibfnamefont{A.~D.} \bibnamefont{Martin}},
  \bibinfo{author}{\bibfnamefont{W.~J.} \bibnamefont{Stirling}},
  \bibinfo{author}{\bibfnamefont{R.~S.} \bibnamefont{Thorne}},
  \bibnamefont{and} \bibinfo{author}{\bibfnamefont{G.}~\bibnamefont{Watt}},
  \bibinfo{journal}{Eur. Phys. J.} \textbf{\bibinfo{volume}{C63}},
  \bibinfo{pages}{189} (\bibinfo{year}{2009}), \eprint{0901.0002}\relax
\mciteBstWouldAddEndPuncttrue
\mciteSetBstMidEndSepPunct{\mcitedefaultmidpunct}
{\mcitedefaultendpunct}{\mcitedefaultseppunct}\relax
\EndOfBibitem
\end{mcitethebibliography}


\end{document}